\documentclass[runningheads,a4paper]{llncs}

\usepackage{amssymb}
\usepackage{amsmath}
\usepackage{graphicx}
\usepackage{url}
\usepackage{xspace}
\usepackage[defblank]{paralist}

\urldef{\mailsc}\path|fariz.darari@stud-inf.unibz.it, {razniewski, nutt}@inf.unibz.it|    
\newcommand{\keywords}[1]{\par\addvspace\baselineskip
\noindent\keywordname\enspace\ignorespaces#1}

\usepackage{epstopdf}
\usepackage{hyphenat}
\usepackage{stmaryrd}
\usepackage{macros_ld}

\begin{document}

\mainmatter  

\title{Bridging the Semantic Gap between RDF and SPARQL using Completeness Statements [Extended Version]}

\titlerunning{Bridging the Semantic Gap between RDF and SPARQL}

\author{Fariz Darari \and Simon Razniewski \and Werner Nutt}
\authorrunning{}
\institute{Faculty of Computer Science, Free University of Bozen-Bolzano, Italy\\
\mailsc\\}

\toctitle{Lecture Notes in Computer Science}
\tocauthor{Authors' Instructions}
\maketitle

\begin{abstract}
RDF data is often treated as incomplete, following the Open-World Assumption. On the other hand, SPARQL, the standard query language over RDF, usually follows the Closed-World Assumption, assuming RDF data to be complete. 
This gives rise to a semantic gap between RDF and SPARQL.
In this paper, we address how to close the semantic gap between RDF and SPARQL in terms of certain answers and possible answers using completeness statements.
This paper is an extended version with proofs of a poster paper~\cite{posterpaper} at ISWC 2014.
\end{abstract}


\section{Introduction}

Due to its open and incremental nature, data on the Semantic Web is generally incomplete.
This can be formalized using the Open-World Assumption (OWA)\cite{reiter1978}.
SPARQL, on the other hand, interprets RDF data under closed-world semantics.
While for positive queries, this does not create problems~\cite{DBLP:conf/pods/ArenasP11},
SPARQL queries with negation make only sense under closed-world semantics.
This ambiguous interpretation of RDF data poses the question how to determine which semantics is appropriate in a given situation.

As a particular example,
suppose we want to retrieve all Oscar winners that have tattoos.
For obvious reasons, no sources on the Semantic Web contain complete information about this topic and hence the Open-World Assumption applies. On the other hand, suppose we want to retrieve all Oscar winners. Here, an RDF version of IMDb\footnote{\url{http://www.imdb.com/oscars/}} would contain complete data and hence one may intuitively apply closed-world reasoning.

In~\cite{DBLP:conf/semweb/DarariNPR13}, completeness statements were introduced, which are metadata that allow one to specify that the CWA applies to parts of a data source.
We argue that these statements can be used to understand the meaning of query results in terms of certain and possible answers~\cite{DBLP:journals/tcs/AbiteboulKG91},
which is especially interesting for queries with negation: 
Suppose an RDF data source has the completeness statement ``complete for all Oscar winners''.
The result of a SPARQL query for Oscar winners contains not only all certain but also all possible answers. The result of a SPARQL query for Oscar winners with tattoos contains only all certain answers. Moreover, the result of a SPARQL query for people with tattoos that did not win an Oscar contains all certain answers, but not all possible answers. 
The result of a SPARQL query for Oscar winners not having tattoos contains all possible answers but none of the answers is certain. 

In this paper, we discuss
how to assess the relationship between certain answers, possible answers and the results retrieved by SPARQL queries in the presence of completeness statements.

\newcommand{\hazanavicius}{Hazanavicius}

\section{Formalization}

\subsubsection{SPARQL Queries.}

A basic graph pattern (BGP) is a set of triple patterns~\cite{W3C13b:sparql}. In this work, we do not consider blank nodes.
We define a \emph{graph pattern} inductively as follows:
(1) a BGP $P$ is a graph pattern;
(2) for a BGP $P$, a \notExists\ pattern $\neg{P}$ is a graph pattern;
(3) for graph patterns $P_1$ and $P_2$,
$P_1 \, \AND \, P_2$ is a graph pattern.
Any graph pattern $P$ can be equivalently written as a conjunction of a BGP (called the positive part and denoted as $P^\positive$)
and several \notExists\ patterns $\set{\neg P_1, \ldots, \neg P_n}$ (referred to as the negative part and denoted as $P^\negative$).
A query with negation has the form $Q = (W, P)$
where $P$ is a graph pattern and $W \subseteq \var{P^\positive}$ is the set of distinguished variables.
%
%
%
%
%
%
The evaluation of a graph pattern $P$ over a graph $G$ is defined in~\cite{W3C13b:sparql}:
\begin{center}
\vspace{-0.4em}
$\eval{P^\positive \, \AND \, \neg P_{1} \, \AND \, \ldots \, \AND \, \neg P_{n}}{G} =$
%
%
$\set{\mu \mid \mu \in \eval{P^\positive}{G} \wedge \forall i \, . \, \eval{\mu(P_{i})}{G} = \emptyset}$
\vspace{-0.4em}
\end{center}
The result of evaluating $(W,P)$ over a graph $G$ is the restriction of $\eval{P}{G}$ to $W$.
This fragment of SPARQL queries is safe, that is, for a query with negation $Q$ and a graph $G$,
the evaluation $\eval{Q}{G}$ returns finite answers.
We assume all queries to be consistent, that is, there is a graph $G$ where $\eval{Q}{G} \neq \emptyset$.

\subsubsection{Completeness Statements.}

%
Formally, a \emph{completeness statement} $C$ is of the form $\complDuo{P_1}{P_2}$ where $P_1$, called the pattern, and $P_2$, called the condition, are BGPs. Intuitively, such a statement expresses that the source contains all instantiations of $P_1$ that satisfy condition $P_2$ (e.g.,\ all Golden Globe winners that won an Oscar).
%
%
%

In line with the Open-World Assumption of RDF,
for a graph $G$,
we call any graph $G'$ such that $G' \supseteq G$ an \emph{interpretation} of $G$~\cite{DBLP:conf/pods/ArenasP11}. 
We associate to a completeness statement $C$ the \CONSTRUCT\ query $Q_C = \buildCONSTRUCT{P_1}{P_1 \, \AND \, P_2}$.
A pair $(G,G')$ of a graph and one of its interpretations
\emph{satisfies} a completeness statement $C$, written $(G,G') \models C$ if $\eval{Q_C}{G'} \subseteq G$ holds.
It satisfies a set $\C$ of completeness statements, written $(G,G') \models \C$ if it satisfies every element in $\C$. 
A set of completeness statements $\C$ \emph{entails} a statement $C$, written $\C \models C$, if for all pairs $(G, G')$ of a graph $G$ and an interpretation $G'$ of $G$ such that $(G, G') \models \C$,
it is the case that $(G, G') \models C$.


Completeness statements restrict the set of interpretations:

\begin{definition}[Valid Interpretation with Completeness Statements]
Let $G$ be a graph and $\C$ be a set of completeness statements. 
An interpretation $G' \supseteq G$ is valid w.r.t.\ $\C$ iff
$(G, G') \models \C$.
We write the set of all valid interpretations of $G$ w.r.t.\ $\C$ as $\extDuo{G}{\C}$.
\end{definition}

%

%


\noindent
\begin{definition}[Certain and Possible Answers with Completeness Statements]
Let $Q$ be a query, $G$ be a graph and $\C$ be a set of completeness statements.
Then the certain and possible answers of $Q$ over $G$ w.r.t.\ $\C$ are
$\caTrio{Q}{G}{\C} = \bigcap_{G' \in \extDuo{G}{\C}} \eval{Q}{G'}$
and
$\paTrio{Q}{G}{\C} = \bigcup_{G' \in \extDuo{G}{\C}} \eval{Q}{G'}$, respectively.

\end{definition}

\noindent
If $\C$ is empty, then $\paTrio{Q}{G}{\C}$ and $\caTrio{Q}{G}{\C}$ correspond to the classical certain and possible answers~\cite{DBLP:journals/tcs/AbiteboulKG91}. With completeness statements, some query results may correspond to the certain and possible answers while others may not.

\begin{example}[Possible Answers]
Consider again the query asking for people that have tattoos. Since this is private information we have no knowledge how complete our data source is, and hence the set of possible answers is (nearly) infinite: Anyone not explicitly stated to have tattoos might still have tattoos.

On the other hand, consider the query for people who won an Oscar. It is relatively easy for a data source to be complete on this topic, by comparing: e.g.,\ to the Internet Movie Database (IMDb).
If a data source is complete for all Oscar winners, then there are no further possible answers.
\end{example}

\noindent
The reasoning for queries with negation is more complex.
By~\cite{DBLP:conf/pods/ArenasP11}, under the OWA, for a monotonic query $Q$ and a graph $G$ the certain answers are $\eval{Q}{G}$, while for queries with negation, the certain answers are empty. With completeness statements, the certain answers of a query with negation can be non-empty.

\begin{example}[Certain Answers]
Consider first a query for people that won an Oscar but no Golden Globe. If a data source is complete both for Oscar winners and Golden Globe winners, then the query result contains all possible and all certain answers.

On the other hand, a query for people with tattoos that did not win an Oscar would only return certain answers, but not all possible answers, because there are probably many more people with tattoos that did not win an Oscar.

The result of a query for people that won an Oscar and do not have tattoos contains all possible answers but no certain answers, because we do not know for certain which Oscar winners have tattoos and which do not.
\end{example}


We next define for each query some completeness statements that allow one to capture the crucial information for getting certain or possible answers.
Knowing about the crucial statements helps in data acquisition
identify
which data is needed in order to achieve desired answer semantics.

\newcommand{\crucpos}{C^+_Q}
\newcommand{\crucneg}{\C^{-}_Q}

\begin{definition}[Crucial Statements]
For a query $Q = (W, P)$, the \emph{positive crucial statement} of $Q$, denoted as $\crucpos$, is the 
statement 
$\complDuo{P^\positive}{\true}$. 
The \emph{set of negative crucial statements} of $Q$, denoted as $\crucneg$, is the set 
$\set{\complDuo{P_1}{P^\positive}, \linebreak \ldots, \complDuo{P_n}{P^\positive}}$,
 where $P_1, \ldots, P_n$ are  from the negative part $P^\negative$ of $Q$.
\end{definition}

The next theorems show that the crucial statements can be used to infer relationships between certain answers, possible answers and SPARQL query results.


\begin{theorem}[Bounded Possible Answers]
\label{th:bounded}
Let $\C$ be a set of completeness statements and $Q$ be a positive query.
Then
\[ \C \models \crucpos \mbox{\qquad implies\ \qquad}  \mbox{for all graphs $G$,\ \ }\paTrio{Q}{G}{\C} = \caTrio{Q}{G}{\C} \  (= \eval{Q}{G}). \]
\end{theorem}


\newcommand{\transfer}[1] {T_\C(#1)}
\newcommand{\prot} {\tilde{P}}

\noindent
\textbf{Proof.} ($\Rightarrow$) Assume $\C \models \crucpos$ holds.
For all graphs $G$,
it holds that $\caTrio{Q}{G}{\C} = \bigcap_{G' \in \extDuo{G}{\C}} \eval{Q}{G'} = \eval{Q}{G}$ since $G \in \extDuo{G}{\C}$ and the query $Q$ is positive (and thus monotonic). 
We will now prove that $\paTrio{Q}{G}{\C} = \eval{Q}{G}$.
By definition, 
it holds that $\paTrio{Q}{G}{\C} \supseteq \eval{Q}{G}$.
It is left to prove that $\paTrio{Q}{G}{\C} \subseteq \eval{Q}{G}$.

Since $Q$ is positive, it has the form $Q = (W, P)$ where $P$ is a BGP.
As $\paTrio{Q}{G}{\C} = \bigcup_{G' \in \extDuo{G}{\C}} \eval{Q}{G'}$,
we have to show that
$\eval{Q}{G'} \subseteq \eval{Q}{G}$ for all $G' \in \extDuo{G}{\C}$, that is,
$\set{\mu\vert_W \mid \mu (P) \subseteq G'} \subseteq \set{\mu\vert_W \mid \mu (P) \subseteq G}$.
Let us fix a graph $G$ and an interpretation $G' \in \extDuo{G}{\C}$.
By definition, it holds that $(G, G') \models \C$.
As we have assumed that $\C \models \crucpos$,
this implies $(G, G') \models \crucpos$.
In our case,
$\crucpos = \complDuo{P}{\true}$ and $Q_{\crucpos} = \buildCONSTRUCT{P}{P}$.
Hence, since $(G, G') \models \crucpos$, we have that $\eval{Q_{\crucpos}}{G'} \subseteq G$
and therefore $\bigcup\set{\mu(P) \mid \mu(P) \subseteq G'} \subseteq G$.
In other words,
we have for all mappings $\mu$ that $\mu(P) \subseteq G'$ implies $\mu(P) \subseteq G$.
This immediately implies $\set{\mu\vert_W \mid \mu (P) \subseteq G'} \subseteq \set{\mu\vert_W \mid \mu (P) \subseteq G}$.
\hfill$\square$
\vspace{0.5em}

\noindent
While the equality $\eval{Q}{G} = \caTrio{Q}{G}{\C}$ always holds, the new insight is that $\eval{Q}{G} = \paTrio{Q}{G}{\C}$.
This means the query results cannot miss any information w.r.t.\ reality.

\begin{theorem}[Queries with Negation]
\label{th:negation}
Let $\C$ be a set of completeness statements and $Q$ be a query.
Then 
\begin{compactenum}
	\item $\C \models \crucneg$ \ \ implies \ \ \ for all graphs $G$, $\caTrio{Q}{G}{\C} = \eval{Q}{G}$;
	\item $\C \models \crucpos \wedge  \crucneg$ \ \ \  implies \ \ for all graphs $G$, $\paTrio{Q}{G}{\C} = \caTrio{Q}{G}{\C} = \eval{Q}{G}$.
\end{compactenum}
\end{theorem}


\noindent
\textbf{Proof (Claim 1).} 
Assume $\C \models  \crucneg$,
which means $\C \models \complDuo{P_1}{P^\positive} \, \wedge \, \cdots \, \wedge \, \complDuo{P_n}{P^\positive}$. Let $G$ be a graph. We want to show that $\caTrio{Q}{G}{\C} = \eval{Q}{G}$.

The inclusion ``$\subseteq$'' holds because
$\caTrio{Q}{G}{\C} = \bigcap_{G' \in \extDuo{G}{\C}} \eval{Q}{G'} \subseteq \eval{Q}{G}$, since $G \in \extDuo{G}{\C}$.

As for the inclusion ``$\supseteq$'', suppose there is a mapping $\mu \in \eval{Q}{G}$.
By the query semantics, there must be a mapping $\mu'$ such that $\mu \subseteq \mu'$
where $\mu' \in \eval{P}{G}$.
Consequently, it must hold that
$\mu' \in \eval{P^\positive}{G}$
and for any \notExists\ pattern $\neg P_i$ in $P$,
it is the case $\eval{\mu'(P_i)}{G} = \emptyset$.
We would like to prove that for any valid interpretation $G' \in \extDuo{G}{\C}$,
it holds
$\mu' \in \eval{P}{G'}$. Note that this implies $\mu \in \eval{Q}{G'}$.

Let us fix an interpretation $G' \in \extDuo{G}{\C}$. We would like to prove that $\mu' \in \eval{P}{G'}$.
Since $G \subseteq G'$ and the pattern $P^\positive$ is positive,
it is the case that $\mu' \in \eval{P^\positive}{G'}$.
It is now left to prove that for any \notExists\ pattern $\neg P_i$ in $P$,
we have that $\eval{\mu'(P_i)}{G'} = \emptyset$.

Let us take an arbitrary $P_i$.
We would like to prove that $\eval{\mu'(P_i)}{G'} = \emptyset$.
We have assumed that $\C \models  \complDuo{P_i}{P^\positive}$.
As $G' \in \extDuo{G}{\C}$,
this means $(G, G') \models \complDuo{P_i}{P^\positive}$ and therefore, $\eval{Q_{\complDuo{P_i}{P^\positive}}}{G'} \subseteq G$, or in other words,
$\eval{\buildCONSTRUCT{P_i}{P^\positive \; \AND \; P_i}}{G'} \subseteq G$.
Thus, it must hold that
$\bigcup\set{\alpha(P_i) \mid \alpha \in \eval{P^\positive \; \AND \; P_i}{G'}} \subseteq G$.
By assumption, $\mu' \in \eval{P^\positive}{G}$ (and thus $\mu' \in \eval{P^\positive}{G'}$) and $\eval{\mu'(P_i)}{G} = \emptyset$ also hold.
Thus, there is no mapping $\nu$ such that $\nu \in \eval{P^\positive \; \AND \; P_i}{G'}$ where $\mu' \subseteq \nu$.
But then, we already know that
$\mu' \in \eval{P^\positive}{G'}$.
Consequently, $\eval{\mu'(P_i)}{G'} = \emptyset$.

\vspace{0.7em}

\noindent
\textbf{Proof (Claim 2).}
Assume $\C \models \crucpos \wedge  \crucneg$. 
Let $G$ be a graph. We would like to prove that $\caTrio{Q}{G}{\C} = \paTrio{Q}{G}{\C} = \eval{Q}{G}$. By Claim 1 of this theorem,
$\caTrio{Q}{G}{\C} = \eval{Q}{G}$,
since $\C \models  \crucneg$.

It is now left to prove that $\paTrio{Q}{G}{\C} = \eval{Q}{G}$. By the definition of possible answers,
it holds that $\paTrio{Q}{G}{\C} \supseteq \eval{Q}{G}$ since $G \in \extDuo{G}{\C}$.
We want to prove that $\paTrio{Q}{G}{\C} \subseteq \eval{Q}{G}$.
This means that for any $G' \in \extDuo{G}{\C}$, it is the case that if there is a mapping $\mu \in \eval{Q}{G'}$ then $\mu \in \eval{Q}{G}$.
Let us take such a mapping $\mu \in \eval{Q}{G'}$.
Recall that $Q$ has the form $(W, P^\positive \, \AND \, \neg P_{1} \, \AND \, \ldots \, \AND \, \neg P_{n})$. Therefore, by $\mu \in \eval{Q}{G'}$, there must exist a mapping $\mu' \supseteq \mu$ such that $\mu' \in \eval{P}{G'}$. This means that $\mu' \in \eval{P^\positive}{G'}$ and for all $P_{i}$, it is the case that $\eval{\mu'(P_{i})}{G'} = \emptyset$.
By Theorem \ref{th:bounded}, it is the case that $\mu' \in \eval{P^\positive}{G}$ since $\mu' \in \eval{P^\positive}{G'}$ and the assumption that $\C \models \crucpos$.
By definition, it holds that $G' \supseteq G$, thus it also holds that
for all $P_{i}$, it is the case that $\eval{\mu'(P_{i})}{G} = \emptyset$.
Consequently, $\mu' \in \eval{P}{G}$ and thus, $\mu \in \eval{Q}{G}$.
\hfill$\square$


\vspace{0.5em}

The first item means that if $\C \models \crucneg$,
then every answer returned by the query is a certain answer.
The second item means that if additionally $\C \models \crucpos$, then there also cannot be any other
possible answers
than those returned by $\eval{Q}{G}$. The completeness statement entailment problems can be solved using standard query containment techniques~\cite{RazniewskiN11:completeness}.

\vspace{-0.5em}

\section{Discussion}

We have shown that in the presence of completeness statements,
the semantics of SPARQL may correspond to the certain answer or possible answer semantics.
Our work is based on the observation that parts of data on the Semantic Web are actually complete.
%
In future research,
we would like to consider explicit negative RDF knowledge and completeness statements over it as an alternative
for getting certain and possible answers.
In this work, we assume all information contained in a graph is correct. An interesting case is when this assumption does not hold in general. We would like to investigate correctness statements as the dual of completeness statements.
A further study is also needed for an effective way of maintenance of completeness statements to cope with information changes.
One possible way is to add timestamps to completeness statements.

\vspace{-0.5em}

\bibliographystyle{unsrt}
\bibliography{references}

\end{document}